\documentclass[12pt,a4paper,final]{iopart}

\usepackage{iopams}
\usepackage{graphicx}
\usepackage[breaklinks=true,colorlinks=true,linkcolor=blue,urlcolor=blue,citecolor=red]{hyperref}
\usepackage{verbatim}
\usepackage{cite}
\usepackage{url}
\usepackage{multirow}
\usepackage{color}

\begin{document}

\title[Pad\'e approximant to transcendental equations]{A Pad\'e approximant approach to two kinds of transcendental equations with applications in physics}

\author{Qiang Luo$^{1}$, Zhidan Wang$^{2}$ and Jiurong Han$^{3}$}

\address{$^1$ Department of Physics, Renmin University of China, Beijing, 100872, China\\
         $^2$ School of Mathematical Science, Yangzhou University, Yangzhou, Jiangsu, 225002, China\\
         $^3$ School of Physical Science and Technology, Yangzhou University, Yangzhou, Jiangsu, 225002, China}
\ead{qiangluo@ruc.edu.cn}

\begin{abstract}
In this paper, we obtain the analytical solutions of two kinds of transcendental equations with numerous applications in college physics by means of Lagrange inversion theorem, and rewrite them in the form of ratio of rational polynomials by second order Pad\'e approximant afterwards from a practical and instructional perspective. Our method is illustrated in a pedagogical manner for the purpose that students at the undergraduate level will be beneficial. The approximate formulas introduced in the paper can be applied to abundant examples in physics textbooks, such as Fraunhofer single slit diffraction, Wien's displacement law and Schr\"odinger equation with single or double $\delta$ potential. These formulas, consequently, can reach considerable accuracies according to the numerical results, therefore they promise to act as valuable ingredients in the standard teaching curriculum.

\vbox{}

\end{abstract}
\section{Introduction}
Transcendental equations of certain types frequently emerge in seemingly unrelated branches of physics, and the roots of these equations appear in a great deal of applications. Furthermore, the ascertainment of zeros of equations is a problem commonly encountered in a broad spectrum of scientific applications. A wide variety of root finding algorithms\cite{url-rootfinding} are available to approximate the solutions to any desired degree of accuracy though, exact analytical solutions to physical problems, which provide better insights into the physical significances of associated parameters than purely numerical solutions, are always desirable and preferable.

The approximation formula, however, plays an unique role in teaching physics since it fills the gap between analytical approach and numerical solution. Frustratedly, many useful approximation techniques, aiming to solve special physical problems, are not really familiar to college students. Quite recently, Kevin Rapedius presented complex resonance states (or Siegert states) that describe the tunnelling decay of a trapped quantum particle by means of Siegert approximation method\cite{approSiegert}. Augusto Bel\'endez \textit{et al} obtained a simple but highly accurate approximate expression for the period of a simple pendulum at the aid of Kidd-Fogg approximate formula\cite{approTaylor}. In this paper, we intend to draw attention to the Lagrange inversion theorem\cite{Lagrangeinv,url-Lagrange} and Pad\'e approximant\cite{padeLagevin} by two kinds of transcendental equations that have made their appearances in a variety of applications. The first kind of equation is
\numparts\label{tanxcotx}
\begin{eqnarray}
\tan x= \kappa x\label{tanx},\\
\cot x= \kappa x\label{cotx},
\end{eqnarray}
\endnumparts
which, splendidly, can be used to find the correlation to the spring oscillations due to the non-ignorable mass of spring\cite{springAJP}, to determine the positions of maxima of Fraunhofer single slit diffraction\cite{diffraction} and the eigenvalues of infinite square potential well with a residual $\delta$-function interaction\cite{doubledelta,chinese}. The second one is
\begin{equation}\label{LambertW}
W(x)e^{W(x)}=x
\end{equation}
where $W(x)$ is the celebrated Lambert $W$ function\cite{Winitzki}, which has undergone a renewed interest during the past three decades and owns a diversity of multidisciplinary applications. Numerous quantities, such as Wien's displacement constant\cite{wienpeak}, can be expressed in closed form in terms of the Lambert $W$ function. Here we refer the motivated readers to \cite{lambertCS,lambertQS} and references therein for more information.

The main purpose of this paper is to provide a pedagogical derivation of the analytical solutions of transcendental equations \eref{tanxcotx} and \eref{LambertW} with the help of Lagrange inversion theorem, and bother Pad\'e approximant subsequently to express them in element forms because of the cumbersome terms of Taylor series expansions. In comparison to other approaches, our formulas are rather transparent and forward, without losing the precision apparently. The mathematical tools are explained briefly in section \ref{sec2}, while the formulas are given afterwards. In section \ref{sec3} and \ref{sec4} we intend to inspire undergraduate students by some typical examples mentioned above after the derivation of the approximate formulas. Section \ref{sec5} is devoted to our conclusions.

\section{Mathematical preliminary: methods and formulas}\label{sec2}
The Lagrange inversion theorem\cite{Lagrangeinv} is a remarkable tool famous for its ability to give explicit formulas where other approaches run into stone walls. Lagrange-B\"urmann formula\cite{url-Lagrange}, a special case of Lagrange inversion theorem, has found many applications, such as evaluating roots of certain transcendental equations and obtaining expansions of a function in powers of a related but different function.

Let the function $f(z)$ be analytic in some neighborhood of the original point $z=0$ of the complex plane with $f(0)\neq0$ and satisfy the equation\cite{Lagrangeinv}
\begin{equation}\label{funcw}
w=\frac{z}{f(z)}.
\end{equation}
There exists two positive numbers $a$ and $b$ such that for $\vert w\vert<a$ the equation\eref{funcw} has just one solution in the domain $\vert z\vert<b$. Lagrange-B\"urmann formula tells us that the unique solution is an analytical function of $w$ with\cite{url-Lagrange}
\begin{equation}\label{BurmannFormula}
z=\sum_{n=1}^{\infty}\frac{w^n}{n!}\Big[\frac{d^{n-1}}{dz^{n-1}}\Big(f(z)\Big)^n\Big]_{z=0}.
\end{equation}

In contract to Taylor expansion, Pad\'e approximat\cite{padeLagevin} has abundant applications in physics because of its fast convergence speed and elegant form. Pad\'e approximant is a type of rational fraction approximation to the value of a function. This structure of approximant enables an effective reconstruction of function's singularities over the whole range using its series expansion obtained for small values of its variable. The $[p,q]$ Pad\'e approximant denotes a fraction of polynomial $P_p(x)=\sum_np_nx^n$ of degree at most $p$, and polynomial $Q_q(x)=\sum_nq_nx^n$ of degree at most $q$
\begin{equation}\label{padepq}
[p,q]\equiv \frac{P_p(x)}{Q_q(x)}.
\end{equation}
The fraction consisting of polynomials $P_p(x)$ and $Q_q(x)$ which approximates function $f(x)=\sum_nf_nx^n$ is determined by the equation\cite{padeLagevin}
\begin{equation}\label{padepq}
f(x) - \frac{P_p(x)}{Q_q(x)}=\mathcal{O}(x^{p+q+1})
\end{equation}
where the symbol $\mathcal{O}(x^{n})$ stands for the value of the order $x^{n}$. $q_0=1$ is always assumed for convenience. It is known that in many cases a higher accuracy of approximation is achieved for fraction of polynomials of identical degree, thus the degrees of both numerator and denominator are set to be 2 hereafter.
\section{Selected applications of transcendental equation \eref{tanxcotx}}\label{sec3}
\subsection{Pad\'e approximant to \eref{tanxcotx}}
We plan to hunt for approximate formulas relating to \eref{tanxcotx} by the Lagrange inversion theorem and Pad\'e approximant in the current section, while the approximate formula of \eref{LambertW} will be provided in the next section.

The analytical solutions to \eref{tanxcotx} are absent until now. From the perspective of graphical method, the solution to \eref{tanx}(or \eref{cotx}) is equivalent to the solution of the pair of equations $y=\kappa x$ and tangent function $y=\tan x$(or cotangent function $y=\cot x$). The discussion will be limited to non-negative solutions because physical quantities involved in current paper only relate to positive roots. Suppose that the zero in the domain $\big[0,\frac{1}{2}\pi\big)$ is denoted by $x_0^{\pm}$ while the zero in the domain $\big[(n-\frac{1}{2})\pi,(n+\frac{1}{2})\pi\big)$ is denoted by $x_n^{\pm}$ with integer $n\geq1$ for \eref{tanx} and \eref{cotx} respectively.

Let us concentrate on the first zero of \eref{tanxcotx} firstly. The first zero of \eref{tanx} is trivial(i.e. $x_0^{+}=0$) for any $\kappa$, while situation is far more complicated for \eref{cotx}. The $[2,2]$ Pad\'e approximant yields\cite{arxiv-2}
\begin{equation}\label{cotpadeapproxbig}
x_0^{-} \approx \frac{1}{\sqrt{\kappa}}\frac{1+\frac{1291}{4044}\kappa^{-1}+\frac{103}{5593}\kappa^{-2}}{1+\frac{655}{1348}\kappa^{-1}+\frac{255}{3704}\kappa^{-2}}
\approx \frac{1}{\sqrt{\kappa}}\frac{\kappa^{2}+0.3192\kappa+0.0184}{\kappa^{2}+0.4859\kappa+0.0688}
\end{equation}
for big $\kappa$, while for small $\vert\kappa\vert$ it is
\begin{equation}\label{cotpadeapproxsmall}
x_0^{-} \approx \frac{\pi}{2}\frac{1+2\kappa+\frac{\pi^2}{12}\kappa^2}{1+3\kappa+(2+\frac{\pi^2}{12})\kappa^2}.
\end{equation}

We pass now to consider the other roots. By means of Lagrange-B\"urmann formula \eref{BurmannFormula}, we arrive at(see appendix A for detail)
\numparts\label{ctanseries}
\begin{eqnarray}
x_n^{+}=\alpha_n\phi_{-}\big(1/\alpha_n\big)\label{tanseries},\\
x_n^{-}=\beta_n\phi_{+}\big(1/\beta_n\big)\label{cotseries},
\end{eqnarray}
\endnumparts
where $\alpha_n=(n+1/2)\pi$, $\beta_n=n\pi$, and $\phi_{\pm}(x)$ are defined as
\begin{equation}\label{fztaylar}
\phi_{\pm}(x)=1\pm\frac{1}{\kappa}x^2-\frac{3\kappa\pm1}{3\kappa^3}x^4+\mathcal{O}(x^5).
\end{equation}
Since $\phi_{\pm}(x)$ are even functions, the $[2,2]$ Pad\'e approximant to \eref{fztaylar} turns out to be
\begin{equation}\label{fzpade}
\phi_{\pm}(x)\approx\frac{(6\kappa\pm1)x^2\pm3\kappa^2}{(3\kappa\pm1)x^2\pm3\kappa^2}.
\end{equation}
Specially, if $\kappa=1$, the solutions to \eref{tanxcotx} are
\numparts\label{ctanpadeapprox}
\begin{eqnarray}
x_n^{+} \approx \alpha_n\frac{3\alpha_n^2-5}{3\alpha_n^2-2}\label{tanpadeapprox},\\
x_n^{-} \approx \beta_n\frac{3\beta_n^2+7}{3\beta_n^2+4}\label{cotpadeapprox}.
\end{eqnarray}
\endnumparts

We mention here that Frankel\cite{Frankel} once obtained a fairly accurate approximation solution to \eref{tanx} in terms of the Taylor series expansion of $\arctan x$. He reached at\cite{Frankel}
\begin{equation}\label{tanFrankel}
x_n^{+} \approx \alpha_n-\big(1+\frac{1}{\alpha_n^2}\big)\mbox{arccot}\alpha_n.
\end{equation}
We note that \eref{tanFrankel} is sightly different from the original version derived by Frankel himself because of the fact that $\arctan x+\mbox{arccot} x={\pi}/{2}$ if $x>0$. Motivated by his inspiring thoughts, we manage to find the counterpart formula for \eref{cotx}, namely
\begin{equation}\label{cotFrankel}
x_n^{-} \approx \beta_n+\frac{1+\beta_n^2}{2+\beta_n^2}\mbox{arccot}\beta_n.
\end{equation}

Detailed comparisons among our formulas and Frankel's formula will be made to end this subsection. \Tref{tabtanx} and \tref{tabcotx} show the accurate values $x_n^{\pm}$ and the absolute errors calculated by (9), (12), \eref{tanFrankel} and \eref{cotFrankel} with $\kappa=1$.
\begin{table}[!h]
\centering
\caption{We present the first 10 non-trivial roots of \eref{tanx} with $\kappa=1$. It can be distinguished from the column of ${x_n^{+}}/{\pi}$ that $x_n^{+}$ approaches to $(n+1/2)\pi$ as $n\rightarrow\infty$. The errors of three different formulas, which are in the unit of $10^{-3}$, are shown in the last three columns in order.}\label{tabtanx}
\begin{tabular}{|*{10}{c|}}
\hline
\multirow{2}*{$n$}
& \multirow{2}*{Exact values}
& \multirow{2}*{$\frac{x_n^{+}}{\pi}$}
& \multicolumn{3}{|c|}{Errors($\times10^{-3}$)} \\\cline{4-6}
& & & Pad\'e   & Frankel & Taylor\\
\hline \hline
 1   &4.49340946 	&1.43029665 	&0.20508427 	&0.45855420 	&0.40225822\\
 \hline
 2   &7.72525184 	&2.45902403 	&0.01474265 	&0.03420796 	&0.02977709\\
 \hline
 3   &10.90412166 	&3.47088972 	&0.00268424 	&0.00629142 	&0.00546478\\
 \hline
 4   &14.06619391 	&4.47740858 	&0.00075749 	&0.00178279 	&0.00154718\\
 \hline
 5   &17.22075527 	&5.48153665 	&0.00027654 	&0.00065221 	&0.00056576\\
 \hline
 6   &20.37130296 	&6.48438713 	&0.00011965 	&0.00028254 	&0.00024503\\
 \hline
 7   &23.51945250 	&7.48647425 	&0.00005841 	&0.00013804 	&0.00011969\\
 \hline
 8   &26.66605426 	&8.48806870 	&0.00003121 	&0.00007378 	&0.00006397\\
 \hline
 9   &29.81159879 	&9.48932662 	&0.00001788 	&0.00004229 	&0.00003667\\
 \hline
10   &32.95638904 	&10.49034444 	&0.00001084 	&0.00002564 	&0.00002222\\
\hline
\end{tabular}
\end{table}
\begin{table}[!h]
\centering
\caption{We present the first 10 non-trivial roots(neglecting $x_0$) of \eref{cotx} with $\kappa=1$. It can be distinguished from the column of ${x_n^{-}}/{n\pi}$ that $x_n^{-}$ approaches to $n\pi$ as $n\rightarrow\infty$. The errors of three different formulas, which are in the unit of $10^{-2}$, are shown in the last three columns in order.}\label{tabcotx}
\begin{tabular}{|*{10}{c|}}
\hline
\multirow{2}*{$n$}
& \multirow{2}*{Exact values}
& \multirow{2}*{$\frac{x_n^{-}}{n\pi}$}
& \multicolumn{3}{|c|}{Errors($\times10^{-2}$)} \\\cline{4-6}
& & & Pad\'e   & Frankel & Taylor\\
\hline \hline
 1   &3.42561846 	&1.09040822 	&-0.36000169 	&-0.18196111 	&-0.87179656\\
 \hline
 2   &6.43729818 	&1.02452783 	&-0.01575732 	&-0.00868217 	&-0.03331845\\
 \hline
 3   &9.52933441 	&1.01109378 	&-0.00222643 	&-0.00124921 	&-0.00458176\\
 \hline
 4   &12.64528722 	&1.00627998 	&-0.00054192 	&-0.00030606 	&-0.00110449\\
 \hline
 5   &15.77128487 	&1.00403118 	&-0.00017970 	&-0.00010180 	&-0.00036460\\
 \hline
 6   &18.90240996 	&1.00280399 	&-0.00007269 	&-0.00004125 	&-0.00014712\\
 \hline
 7   &22.03649673 	&1.00206211 	&-0.00003376 	&-0.00001918 	&-0.00006823\\
 \hline
 8   &25.17244633 	&1.00157982 	&-0.00001736 	&-0.00000987 	&-0.00003505\\
 \hline
 9   &28.30964285 	&1.00124880 	&-0.00000965 	&-0.00000549 	&-0.00001947\\
 \hline
10   &31.44771464 	&1.00101185 	&-0.00000571 	&-0.00000325 	&-0.00001151\\
\hline
\end{tabular}
\end{table}
It can be distinguished that the formulas based on Pad\'e approximant have an advantage over Frankel's formula sightly, but the advantage will be extended with the increase of the order of Pad\'e approximant. Besides, both tables tell us that Pad\'e approximant is around twice as precise as Taylor series expansion, which, of course, suggests the superiority of Pad\'e approxiant.
\subsection{Selected applications of transcendental equation \eref{tanxcotx}}
\subsubsection*{Effect of spring mass on the frequency of oscillator}
The oscillation of a spring-mass system, where a spring is suspended vertically and a mass $m$ is hung from the bottom end of the spring, is a rather classical topic commonly studied theoretically and experimentally in introductory physics courses. The consideration of the correlation to the spring oscillation due to the non-ignorable mass of spring has led to many papers, and the  empirical law that 1/3 the mass of the spring should be added to the mass of the hanging object is frequently-quoted(see reference \cite{springAJP} and references therein). The motion of this celebrated system is governed by\cite{url-springmass}
\begin{equation}\label{definiteproblem}
\left\{
             \begin{array}{lcr}
             u_{tt}-a^2u_{xx}=0, \\
             u\vert_{x=0}=0, \big(u_x+\frac{m}{Ys}u_{tt}\big)\vert_{x=l}=0, \\
             u\vert_{t=0}=\frac{A_0}{l}x,u_t\vert_{t=0}=0,
             \end{array}
\right.
\end{equation}
where $Y,s$ and $\rho$ are the Young's modulus, cross-sectional area and mass density of the spring, and $a=\sqrt{Y/\rho}$ is the wave speed in the spring. The solution to \eref{definiteproblem} is
\begin{equation}\label{definitesolution}
u(x,t)=\sum_{n=1}^{\infty}\frac{\frac{2\rho s}{\sqrt{\rho^2s^2+\lambda_n m^2}}}{l\lambda_n\big(l+\frac{m\rho s}{\rho^2s^2+\lambda_n m^2}\big)}A_0\cos(\sqrt{\lambda_n}at)\sin(\sqrt{\lambda_n}x)
\end{equation}
where $\lambda_n$ satisfies the transcendental equation
\begin{equation}\label{lambdan}
\cot(\sqrt{\lambda}l)=\frac{m}{m_0}\sqrt{\lambda}l.
\end{equation}
Taking into account the Hook's law in the language of Young's modulus, which is defined as the ratio of the stress(force per unit area) along an axis to the strain(ratio of deformation over initial length) along that axis in the range of stress, we obtain the relation $Y=kl/s$ where $k$ is the stiffness of ideal spring, and therewith \eref{lambdan} becomes
\begin{equation}\label{omegan}
\cot\Big(\omega\sqrt{\frac{m_0}{k}}\Big)=\frac{m}{m_0}\Big(\omega\sqrt{\frac{m_0}{k}}\Big)
\end{equation}
where the angular frequency $\omega=a\sqrt{\lambda}$. After a trivial trick, the frequency of the spring turns out to be
\begin{equation}\label{oemgam}
\omega=\sqrt{\frac{k}{m+\xi m_0}}
\end{equation}
where
\begin{equation}\label{equationseries}
\left\{
             \begin{array}{lcr}
             \xi=\frac{1}{\eta^2}-r \\
             \cot(\eta)=r\eta
             \end{array}
\right.
\end{equation}
with $r=m/m_0$. It can be concluded from \eref{definitesolution} that the motion of the spring is the superposition of infinite numbers of simple harmonic vibrations. However, the amplitudes of all the other vibrations are much smaller than the principal vibration's. Many authors, such as Christensen\cite{springAJP}, showed explicitly that the small $r$ case necessarily transitions to the large $r$ case so that the lowest normal mode is always the dominant one. In this occasion, the fundamental frequency of the spring can be calculated by \eref{cotpadeapproxbig} or \eref{cotpadeapproxsmall} subsequently.

Now let us consider two extreme cases. If $r$ is large enough, then $\eta\approx\frac{1}{\sqrt{r}}(1-\frac{1}{6r})$, therefore $\xi_{\min}=\lim\limits_{r\rightarrow\infty}r[(1-\frac{1}{6r})^{-2}-1]=\frac{1}{3}$. On the contrary, $\xi_{\max}=\lim\limits_{r\rightarrow0}[\frac{4}{\pi^2}(1-r)^{-2}-r]=\frac{4}{\pi^2}$ since $\eta\approx\frac{\pi}{2}(1-r)$ when $r$ is a small number.
Specially, if the mass of the spring and hung object are equal, namely $r=1$, one obtain $\eta\approx0.86$ from \eref{cotpadeapproxbig} and therefore find the effective mass coefficient is $\xi\approx0.35$, which is in accordance with the fact that $\frac{1}{3}\leq\xi\leq\frac{4}{\pi^2}$ and satisfies the empirical law mentioned above.
\subsubsection*{Fraunhofer single slit diffraction}
One of the most long-standing problems of classical optics is the various types of diffraction. The theory of Fraunhofer single slit diffraction predicts that the spatial pattern of light intensity on the viewing screen by a light wave passing through a single rectangular-shaped slit is given by\cite{diffraction}
\begin{equation}\label{GuangQiang}
I=I_0\frac{\sin^2u}{u^2}
\end{equation}
where $u=\pi b\sin\theta/\lambda$ and $I_0$ is the light intensity at $\theta=0^{\circ}$. The first and second derivative of $I$ with respect to $u$(or $\theta$ more precisely) are badly needed in order to find the maxima and minima of the diffraction pattern. The minima occur when $u=n\pi$, $n=\pm1,\pm2,\cdots$, while the maxima exist on condition that
\begin{equation}\label{lighttan}
\tan u=u.
\end{equation}
The trivial zero of \eref{lighttan}, i.e. $u=0$ corresponds to the primary maximum of the diffraction pattern, while the non-trivial zeros indicate the secondary maxima. We see that the secondary maxima are not exactly half way between any two adjacent minima, they occur slightly earlier and move toward the center with increasing shift $u$. Figure below shows the variation of the intensity distribution with the distance.
\begin{figure}[!h] 
\centerline{\includegraphics[height=6.0cm]{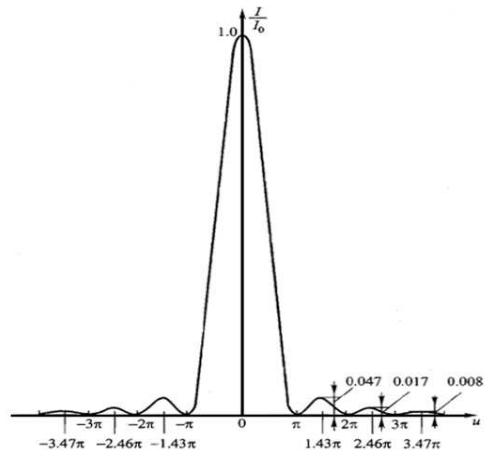}\hspace{4mm}}
\caption{The diffraction pattern of Fraunhofer single slit diffraction. It can be distinguished that intensities of secondary maxima are dramatically less than the principal maximum, and the positions of secondary maxima are not exactly half way between the adjacent minima.}
\end{figure}

Furthermore, intensities of these secondary maxima are much less than primary maximum and fall off rapidly as moving outwards. The relative intensity of the $n$-th secondary maximum to the primary maximum can be determined from \eref{tanpadeapprox} as
\begin{equation}\label{XDGQ}
\frac{I_n}{I_0}=\frac{1}{2}\big(\frac{9}{3\alpha_n^2-2}-\frac{1}{\alpha_n^2}\big)
\end{equation}
where $\alpha_n=(n+1/2)\pi$. For example, the relative intensity of the first three secondary maxima are only $4.7\%$, $1.7\%$ and $0.8\%$ of the principal maximum, respectively.
\subsubsection*{Schr\"odinger equation with single $\delta$ potential}
We shall discuss a problem encountered in many introduction course of quantum mechanics in detail. Now let's consider a microscopic particle confined in an 1-dimensional infinite square well in $0<x<a$, and assume that a $\delta$-function interaction is added in the middle of the interval with constant strength $\gamma$, whose sign indicates the interaction is attractive or repulsive. Therefore, the potential of the particle is\cite{chinese,arxiv-2}
\begin{equation}\label{potential}
V(x)=
   \cases
   {\gamma\delta(x-a/2), &$0<x<a$,\\
   +\infty,             &\mbox{otherwise}}.
\end{equation}

The wavefunctions can be classified according to their symmetries under the transformation $\psi_n(x)=\pm\psi_n(a-x)$ since the Hamiltonian is symmetric with respect to the transformation $x\rightarrow a-x$. Thereby only the even parity eigen-states $\psi_n(x)$, which satisfy the Schr\"odinger equation
\begin{equation}\label{schrodinger}
-\frac{\hbar^2}{2m}\frac{d^2}{dx^2}\psi(x)+\gamma\delta\big(x-\frac{a}{2}\big)\psi(x)=E\psi(x),
\end{equation}
are influenced by the additional interaction. At point $x=a/2$ the wavefunction should be continuous but its derivative makes a jump proportional to the strength of the $\delta$-function interaction, i.e.
\begin{equation}\label{schrodingerjump}
\psi'\big(\frac{a}{2}+\epsilon\big)-\psi'\big(\frac{a}{2}-\epsilon\big)=\frac{2m\gamma}{\hbar^2}\psi\big(\frac{a}{2}\big)
\end{equation}
with $\epsilon\rightarrow0$. In general, the energy $E_n$ relates to the quasi wave-vector $k_n$, that satisfies the energy eigenvalues condition:
\begin{equation}\label{schrodingereig}
\tan\big(\frac{ka}{2}\big)=-\frac{b}{a}\big(\frac{ka}{2}\big)
\end{equation}
where $b\equiv2\hbar^2/m\gamma$ is the characteristic length of $\delta$-potential. The energy are positive if the interaction is repulsive, situation will be more complicated if the interaction is attractive. In the case of $\gamma<0$, ground state energy for instance, will go through a process from positive to negative with increasing intensity of the interaction. However, unlike the other eigenstates, the $E=0$ energy eigenstate of the Schr\"odinger equation is not a sinusoidal function. Instead, the time-independent Schr\"odinger equation simplifies to ${d^2\psi}/{dx^2}=0$ inside the well, and yields $\psi=Ax+B$, where $A$ and $B$ are constants\cite{arxiv-2}. The energy eigenstate has the proper discontinuity in its slope at the middle point of the well such that\cite{chinese}
\begin{equation}\label{psi}
\psi(x)=
   \cases{
   2\sqrt{\frac{3}{a^3}}x,             &$0\leq x<a/2$,\\
   2\sqrt{\frac{3}{a^3}}\big(a-x\big), &$a/2<x\leq a$.
   }
\end{equation}
The combination of \eref{psi} and \eref{schrodingerjump} yields the critical intensity $\gamma_0=-\frac{2\hbar^2}{ma}$, and therefore \eref{schrodingereig} is reduced to
\begin{equation}\label{schrodingereig0}
\tan\big(\frac{ka}{2}\big)=\frac{ka}{2}.
\end{equation}
\Eref{tanpadeapprox}, readily, tells us that the solution of \eref{schrodingereig0} satisies the identity $\big(k_na/2\big)^2-\big(n\pi/2\big)^2\approx-2\big(1+{2}/{3(n\pi)^2}\big)$ with $n=3,5,7,\cdots$ approximately, then the even-parity energy levels of the system considered are
\begin{equation}\label{energydelta}
E_n^{\bf{e}}=\frac{\hbar^2k_n^2}{2m}
   \cases{
   =E_1^{(0)}+\frac{\pi^2}{4}\frac{\gamma_{0}}{a},                            &$n=1$,\\
   \approx E_n^{(0)}+2\Big(1+\frac{2}{3(n\pi)^2}\Big)\frac{\gamma_{0}}{a},    &$n=3,5,7,\cdots$
   }
\end{equation}
where $E_n^{(0)}=\frac{n^2\pi^2\hbar^2}{2ma^2}$ is the energy of one-dimensional infinity square well without $\delta$-potential. The errors of the energy obtained from \eref{energydelta}, which can be inferred from \tref{tabtanx}, are rather small. Furthermore, \eref{energydelta} suggests us that the $\delta$-potential has a more significant influence on ground state energy than the excited state energy since the prefactor $\pi^2/4(\approx2.5)$ for $n=1$ is larger than $2\big(1+{2}/{3(n\pi)^2}\big)\approx2$ that for $n=3,5,7,\cdots$.
\section{Selected applications of transcendental equation \eref{LambertW}}\label{sec4}

\subsection{Pad\'e approximant to \eref{LambertW}}
The Lambert $W$ function is a multivalued function with complex variable though, of special relevance to scientific applications are the solutions when the argument is purely real number. The two real solutions of \eref{LambertW} are the branches $W_0$ and $W_{-1}$ where $W_0$ is the principal branch of the $W$ function.
\begin{figure}[!h] 
\centerline{\includegraphics[width=6.5cm]{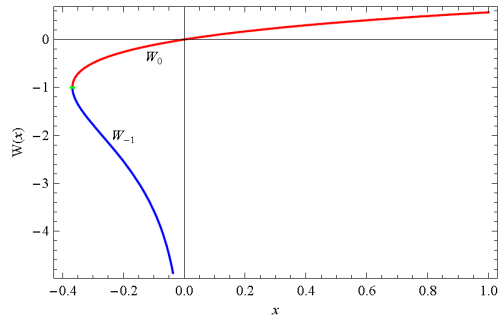}\hspace{4mm}}
\caption{Two real branches of the Lambert $W$ function. Red line: $W_0(x)$ called the principal
branch is defined for $-1/e < x < +\infty$. Blue line: $W_{-1}(x)$ is defined for $-1/e < x< 0$. The
two branches meet at the green point $(-1/e,-1)$\cite{arxiv-1}.}
\end{figure}

The principal branch $W_0$ is analytical at $x=0$ and its Taylor series expansion\cite{lambertCS,lambertQS}
\begin{equation}\label{powLambert}
W_0(x)=\sum_{n=1}^{\infty}\frac{(-n)^{n-1}}{n!}x^n=x\Big(1-x+\frac{3}{2}x^2-\frac{8}{3}x^3+\frac{125}{24}x^4+\mathcal{O}(x^5)\Big)
\end{equation}
can be obtained in light of Lagrange inversion theorem. The $[2,2]$ Pad\'e approximant reads
\begin{equation}\label{pade1a}
W_0^{(\bf{I})}(x)=x\frac{1+\frac{19}{10}x+\frac{17}{60}x^2}{1+\frac{29}{10}x+\frac{101}{60}x^2}+\mathcal{O}(x^5).
\end{equation}
After rounding the fractional coefficients by the special method introduced in \cite{padeLagevin}, \eref{pade1a} is reduced to
\begin{equation}\label{pade1b}
\overline{W}_0^{(\bf{I})}(x)=x\frac{3+6x+x^2}{3+9x+5x^2}+\mathcal{O}(x^5).
\end{equation}
which, of course, is not only more elegant but isn't accomplished at the sacrifice of accuracy. In fact, the substitution $x\rightarrow\ln(1+x)$ of the pre-factor is a good succedaneum since the slope of $W(x)$ around zero is no more than 1. We therefore arrive at
\begin{equation}\label{pade2a}
W_0^{(\bf{II})}(x)=\ln(1+x)\frac{1+\frac{123}{40}x+\frac{21}{10}x^2}{1+\frac{143}{40}x+\frac{713}{240}x^2}+\mathcal{O}(x^5)
\end{equation}
if we utilize the fact that $\ln(1+x)=\sum_n(-1)^{n+1}x^n/n$. Readers can refer to the appendix below for more details. Similar method impels result such that
\begin{equation}\label{pade2b}
\overline{W}_0^{(\bf{II})}(x)=\ln(1+x)\frac{2+6x+4x^2}{2+7x+6x^2}+\mathcal{O}(x^5).
\end{equation}

\begin{figure}[!htb]  
\centerline{\includegraphics[width=7.0cm]{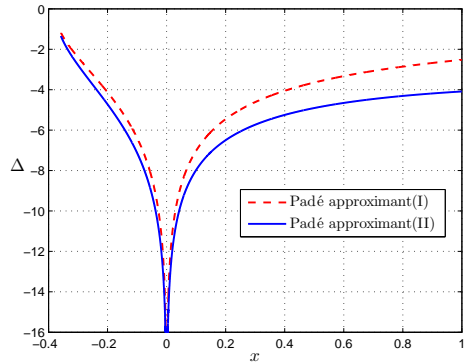}\hspace{4mm}}
\caption{The curves of the logarithm of the absolute errors($\Delta=\lg\big(\vert\frac{W_0^{A}(x)-W_0^{E}(x)}{W_0^{E}(x)}\vert\big)$) regarding to the two different formulas in the interval $-1/e<x<1$ are plotted where $W_0^{A}(x)$ and $W_0^{E}(x)$ are the approximate and exact values of the $W$ function. The dashed line and the solid line, from the top down, represent Pad\'e formula of type \mbox{I} and \mbox{II} respectively.}\label{figError}
\end{figure}

\Fref{figError} shows the precision of the two formulas, i.e. \eref{pade1a} and \eref{pade2a} introduced above. It can be concluded that Pad\'e formula of type \mbox{II} is superior to type \mbox{I} overwhelmingly, especially for positive argument. Therefore, we recommend that \eref{pade2a} is the perfect approximate formula.

Last but not the least, let us focus on an equation relating to Lambert $W$ function. The equation is
\begin{equation}\label{ExponLinear}
e^{-cx}=a(x-b)
\end{equation}
where $a,b,c$ are constants. A great deal of quantities, such as the decay constant of an exponentially decaying process\cite{wienLamEJP} and the time constant of a process subject to a linear resistive force\cite{Winitzki}, satisfy the same equation as shown above. The solution to \eref{ExponLinear} is
\begin{equation}\label{ExponLinearSolution}
x=b+\frac{1}{c}W\big(\frac{c}{a}e^{-cb}\big).
\end{equation}

\subsection{Selected applications of transcendental equation \eref{LambertW}}
\subsubsection*{Schr\"odinger equation with double $\delta$ potential}
Let's extend to the case of a particle in a double delta function well which can be used to describe the behaviour of electronic terms of $\mbox{H}_2^{+}$. Mathematically, the potential is described by\cite{doubledelta}
\begin{equation}\label{potentialdouble}
V(x)=-\gamma\Big(\delta(x+a)+\delta(x-a)\Big)
\end{equation}
where $\gamma,a>0$. We wouldn't repeat the tedious derivation again, instead, we recommend the readers to some professional literatures, such as \cite{doubledelta} for details. Here we just sketch the general conclusions. The potential is an even function, so all the solutions can be expressed as a linear combination of even and odd solutions, namely the so-called even parity and odd parity. The bound energy of the particle, in either situation, can be expressed as
\begin{equation}\label{deltaenergydouble}
E^{\pm}=-\frac{\hbar^2(k^{\pm})^2}{2m}
\end{equation}
where the wavevector $k^{\pm}$ is the solution to equation
\begin{equation}\label{deltapm}
k=\frac{1}{2b}(1\pm e^{-2ka})
\end{equation}
with $b\equiv\frac{\hbar^2}{2m\gamma}$, and the plus and minus sign correspond to the even parity and odd parity solution, respectively. Thanks to \eref{ExponLinearSolution} we manage to obtain the bound state energy
\begin{equation}\label{deltaenergydoubleLambert}
E^{\pm}=-\frac{\hbar^2}{8ma^2}\Big[\frac{a}{b}+W\big(\pm\frac{a}{b}e^{-\frac{a}{b}}\big)\Big]^2.
\end{equation}
Therefore, it can be inferred that the bound state energy always exists for even parity, while for odd parity it only exists on condition that $a>b$. In other words, there're two bound states if the intensity of the interaction $\gamma>\frac{\hbar^2}{2ma}$, otherwise there exists only one bound state. Our approximation formula \eref{pade2a} is sufficient to determine the energy since the argument of the Lambert $W$ function in \eref{deltaenergydoubleLambert} is bounded from $-1/e$ to $1/e$ for both cases. For example, if $a=2b$, the bound state energy are $-0.6148\frac{\hbar^2}{ma^2}$ and $-0.3176\frac{\hbar^2}{ma^2}$ respectively, while the accuracy values presented in \cite{doubledelta} are $-0.614782\frac{\hbar^2}{ma^2}$ and $-0.317454\frac{\hbar^2}{ma^2}$ respectively. The fact that the relative errors are $0.003\%$ and $0.05\%$ or so indicates that our approximate results are in glorious agreement with the exact ones. The exchange energy $\Delta E$ defined as the difference between $E^{+}$ and $E^{-}$ can be determined subsequently.

\subsubsection*{Wien's displacement law}
Planck's seminal work for blackbody radiation, whose result conflicted dramatically with classical mechanics, opened the era of quantum theory. The concept of blackbody radiation, along with the associated Stefan-Boltzmann law and Wien's displacement law, is a crucial pillar of modern physics. The Planck spectral distribution is given by\cite{wienLamEJP}
\begin{equation}\label{Planck}
\rho(\lambda,T)=\frac{8\pi hc}{\lambda^5}\cdot\frac{1}{e^{hc/\lambda kT}-1}
\end{equation}
where $\lambda$ is the wavelength, $T$ is the temperature, $c$ is the speed of light, and $h$ and $k$ are the Planck and Boltzmann constants respectively. Wien's displacement law gives the wavelength at which \eref{Planck} has the maximal intensity. To find the extremum of the specific intensity, it's necessary for us to take the derivative of \eref{Planck} with respect to wavelength $\lambda$ and utilize the often-used unitless variable $x=hc/\lambda kT$. Apparently, we have the fancy equation
\begin{equation}\label{peakequation}
(5-x)e^{x}=5
\end{equation}
whose non-trivial zero is of great interest. In order to solve \eref{peakequation} analytically, several attempts have been made to express the root in integral representation and series representation over the past one century. Siewert and Burniston found that the analytical solution of \eref{LambertW} relates to the canonical solution of the Riemann problem\cite{LambertSiewert}, and Siewert applied it to Wien's displacement law to observe that\cite{wienNoteint}
\begin{equation}\label{peakint1}
x_0=4\exp\Big(-\frac{1}{\pi}\int_0^{\infty}\Big[\arctan\Big(\frac{\pi}{\ln 5-5-t-\ln t}\Big)-\pi\Big]\frac{dt}{t+5}\Big).
\end{equation}
Luck and Stevens, at the meantime, presented another integral representation by virtual of Cauchy's integral theorem and some basic concepts of complex integration\cite{wienSIAM}, i.e.
\begin{equation}\label{peakint2}
x_0=5\frac{\int_0^{2\pi}w(\theta)e^{i3\theta}d\theta}{\int_0^{2\pi}w(\theta)e^{i2\theta}d\theta}
\end{equation}
where $w(\theta)=\frac{1}{5}\frac{1}{(1-e^{i\theta})e^{5e^{i\theta}}-1}$. On the other hand, Andersen\cite{wienNoteseries} and Vial\cite{wienLamEJP} suggested that only the first three terms of the Taylor series expansion in \eref{powLambert} can reach a satisfactory precision.

As a matter of fact, \eref{peakequation} is nothing but \eref{ExponLinear} with $a=1/5$, $b=5$ and $c=1$. Therefore Wien's displacement law can be obtained with an elegant expression for displacement constant $b={hc}/{kx_0}$ where
\begin{equation}\label{x0Lambert}
x_0=5+W_0(-5e^{-5}).
\end{equation}
It can be distinguished from \eref{peakint1}, \eref{peakint2} and \eref{x0Lambert} that the nontrivial zero expressed by Lambert $W$ function pushes the mathematical structure of the law into the most comfortable territory, and the calculation based on it is especially accuracy. Furthermore, our approximation formula \eref{pade2a} gives $x_0=4.965114231797$, which owns ten decimal places precision compared to the accuracy value $x_0=4.965114231744$. The relative error, without doubt, is rather small, and it is of the order $10^{-11}$. What's more, even the simplified version \eref{pade2b} can uncover four decimal places precision.
\section{Conclusion}\label{sec5}
In spite of the fact that the solutions to many transcendental equations, such as \eref{tanxcotx} and \eref{LambertW}, can be expressed analytically in power series or named special functions, but it's not really convenience for instructional purpose in classroom. This paper has provided a detailed illustration of how Lagrange inversion theorem and Pad\'e approximant can be applied to solve transcendental equations analytically, and express the root(s) of those equations in the form of ratio of rational polynomials. Furthermore, our approximation method should be quite powerful in handling with the non-transparent aspect of some rigourous results, which could eventually be expressed in more brief forms, without losing the physical interpretations. Traditional graphical method, however, is not only troublesome but too oversimplified to reach desired accuracy. While all of the drawbacks can be overcome by the built-in algorithms in the softwares such as Matlab and Mathematica, the mathematical softwares tend to give numerical results, or just make the so-called closed-form solutions be lengthy, which are useless in practice. Our method, therefore, is superior in teaching because it is easy to handle and owns satisfactory accuracy. An excellent agreements were achieved between our approximate results and analytical formulas, which prove that those formulas are rather precise in practice, and therefore can act as valuable ingredients in the standard teaching curriculum.
\section*{Acknowledgements}
The authors would like to thank two anonymous referees for helpful suggestions on improving the manuscript. Q. Luo would especially express his appreciation to Professor Q.H. Liu, working at Hunan University, for his selfless help during the manuscript preparation.

\section*{Appendix A}\label{appA}
\setcounter{equation}{0}
\renewcommand{\theequation}{A.\arabic{equation}}
we begin the process of putting the equation\eref{tanx} into the form required by the Lagrange inversion formula firstly. The asymptotic expression $x_n^{+}\sim(n+\frac{1}{2})\pi$ holds as $n\rightarrow\infty$. Let us set $z=w^{-1}-x$ where $w=\alpha_n^{-1}$, then \eref{tanx} is reduced to
\begin{equation}\label{derive}
\tan(x)=\tan\Big(\big(n+\frac{1}{2}\big)\pi-z\Big)=\cot(z)=\kappa(w^{-1}-z).
\end{equation}
Therefore, \eref{derive} can be arranged to \eref{funcw} with
\begin{equation}\label{fztanx}
f_{+}(z)=\frac{z(\cos z+ \kappa z\sin z)}{\kappa\sin z}.
\end{equation}
Similarly, we can get the expression for $f_{-}$. We thus can obtain the approximate formulas to \eref{tanxcotx} according to \eref{BurmannFormula} since $f_{\pm}(z)$ is analytical around $z=0$ and $f(0)=1/\kappa$.
\section*{Appendix B}\label{appB}
\setcounter{equation}{0}
\renewcommand{\theequation}{B.\arabic{equation}}
The calculation of $W_0^{(\bf{II})}(x)$, where the pre-factor is $\ln(1+x)$ rather than $x$, is not really straightforward. To begin with, let us define the auxiliary function
\begin{equation}\label{Mfun}
M(x)=\frac{W(x)}{\ln(1+x)}=\sum_{n=0}^{\infty}a_nx^n
\end{equation}
where $a_n$ are the Taylor expansion coefficients. if we utilize the fact that $\ln(1+x)=\sum_n(-1)^{n+1}x^n/n$ and the equation \eref{powLambert}, we find that
\begin{equation}\label{MfunTaylor}
M(x)=1-\frac{1}{2}x+\frac{11}{12}x^2-\frac{43}{24}x^3+\frac{2651}{720}x^4+\mathcal{O}(x^5)
\end{equation}
after rearranging the coefficients of the same power. We therefore can get the second order Pad\'e approximant of the Lambert $W$ function around zero as is shown in \eref{pade2a} by virtue of the Pad\'e approximant of \eref{MfunTaylor}
\begin{equation}\label{MfunPade}
M(x)=\frac{1+\frac{123}{40}x+\frac{21}{10}x^2}{1+\frac{143}{40}x+\frac{713}{240}x^2}+\mathcal{O}(x^5).
\end{equation}

\section*{References}


\end{document}